\documentstyle[prl,multicol,aps]{revtex}
\begin{document}

\input{epsf.tex}

\title{Interplay between Coherence and Incoherence in Multi-Soliton Complexes}

\author{Andrey A. Sukhorukov and Nail N. Akhmediev}

\address{Optical Sciences Centre,
Research School of Physical Sciences and Engineering,
The Australian National University, Canberra ACT 0200, Australia}
\maketitle

\begin{abstract}

We analyze photo-refractive incoherent soliton beams and their interactions
in Kerr-like nonlinear media. 
The field in each of $M$ incoherently interacting components is calculated
using an integrable set of 
coupled nonlinear Schr\"odinger equations. 
In particular, we obtain a general $N$-soliton solution, describing
propagation of multi-soliton complexes and their collisions.
The analysis shows that the evolution
of such higher-order soliton beams is determined 
by coherent and incoherent contributions from fundamental solitons.
Common features and differences between these internal interactions
are revealed and illustrated by numerical examples.

\end{abstract}

\pacs{PACS number: 42.65Tg, 42.65Jx}

\begin{multicols}{2}
\begin{narrowtext}

One of the most noted discoveries of modern soliton science is that
solitons can be excited by an incandescent light bulb instead of a high
power laser source \cite{Mitchell_Segev}. This produces
"incoherent solitons" \cite{Mitchell_Segev,Incohfirst,newstuff}; they
can exist in photorefractive materials which require amazingly low
powers to observe highly nonlinear phenomena
\cite{pr_solitons1,pr_solitons6,pr_solitons2}. It is also remarkable that,
in certain conditions, incoherent solitons in photorefractive materials
can be studied using coupled nonlinear Schr\"odinger equations
(NLSE) \cite{newstuff,Vysl2}.

In general, coupled nonlinear Schr\"odinger equations (NLSE) 
can be applied to various phenomena.
These include incoherent solitons in photo-refractive materials,
plasma waves in random phase approximation \cite{Hasegawa1}, 
multicomponent Bose-Einstein condensate \cite{Bashkin} and self-confinement
of multimode optical pulses in a glass fiber \cite{Hasegawa2}.
Therefore its solutions are of great interest for theoretical
physicists. In special cases these equations are found to be
integrable \cite{Makh1}. Then, in analogy with single (scalar) NLSE
(when the number of equations, $M$, is $1$) \cite{ZS} and the Manakov case
\cite{Manakov} ($M=2$), the total solution consists of a finite number ($N$)
of solitons and small amplitude radiation waves. The former is defined by
the discrete spectrum of linear \mbox{$(L,A)$} operators \cite{ZS,Manakov} 
and the latter is defined by the continuous spectrum. Most applications deal
with the soliton part of the solution as it contains the most important
features of the problem. Moreover, a localized superposition of
fundamental solitons can be called
``multisoliton complex''. An incoherent soliton is a particular example of
a multisoliton complex \cite{PCS}.

The cases $M=1$ and $M=2$ have been extensively discussed in the
literature \cite{ZS,Manakov}.
On the other hand, results for general $M$ are scarce. The linear 
\mbox{$(L,A)$} operators are important elements for the inverse scattering
technique, which can be considered as a basis for integrability of $M$
coupled NLSEs. Moreover, it has been shown \cite{Nogami} that $N$-soliton
solutions of $M$ coupled NLSE can be found using a simple technique which is
an extension of the theory of reflectionless potentials \cite{Kaw75}.
In recent works \cite{PCS,Ank99} cases when each component has only one
fundamental soliton have been considered. 
It was demonstrated that, in this configuration, the
formation of stationary complexes may be observed, 
and corresponding
solutions for $M = N \leq 4$ were presented in explicit form
\cite{Ank99}.

So far, only the case of complete mutual incoherence of the fundamental 
solitons has been considered. In this case the multisoliton complex
can also be viewed as a self-induced multimode waveguide \cite{PCS}.
The general case, where fundamental solitons in the multisoliton
complex interact both coherently and incoherently, has not been analyzed.
Such interactions may be observed if $N$ is larger than $M$, so that each
component has not less than one fundamental soliton. 
In general, each fundamental soliton can be
"spread out" among several components. We will refer to this effect as
mixed ``polarization'' of fundamental solitons. However, in order to capture
distinctive features of coherent and incoherent soliton interactions, we will 
focus on a special case which is important for incoherent solitons. 
Specifically, we consider a situation where
all the fundamental soliton polarizations are mutually parallel or orthogonal, 
and thus are conserved in collisions \cite{PCS}.
Due to the symmetry of the NLSE with respect to rotations
in functional space, hereafter we assume that each fundamental soliton 
is polarized in one component only. 
It is for this case that we present new explicit $N$-soliton
solutions of $M$ coupled NLSEs, and we discuss the new physics 
which it brings into the theory.

We consider propagation of an incoherent self-trapped beam
in a slow Kerr-like medium and write the set of coupled NLSEs
in the form \cite{newstuff,Vysl2,PCS}:
\begin{equation}  \label{eq:A}
 i \frac{\partial \psi_{m}}{\partial z}
 + \frac{1}{2} \frac{\partial^{2}\psi_{m}}{\partial x^{2}}
 + \delta n(I) \psi_{m} = 0,
\end{equation}
where $\psi_m$ denotes the $m$-th component of the beam,
$z$ is the coordinate along the direction of propagation,
$x$ is the transverse coordinate, and
\begin{equation}  \label{eq:B}
\delta n(I) = \sum\limits_{m=1}^{M}\alpha_m |\psi_{m}|^{2}
\end{equation}
is the change in refractive index profile created by all incoherent
components of the light beam, where the $\alpha_m$ \mbox{$(>0)$} are the 
coefficients representing the strength of the nonlinearity, 
and $M$ is the number of components.

Solutions in the form of multisoliton complexes of Eq.~(\ref{eq:A})
and their collisions can be obtained using the formalism of
\cite{Nogami,Gardner} with some refinements.
First, we introduce
functions $u_j(x,z)$ as solutions of the following set of equations:
\begin{equation} \label{eq:slau}
 \sum_{m=1}^N {D_{jm} u_m} = - e_j.
\end{equation}
where $N$ is a total number of fundamental solitons,
$e_j = \chi_j \exp \left( k_j \bar{x}_j + i k_j^2 \bar{z}_j / 2 \right)$,
$\bar{x}_j = x - x_j$ and
$\bar{z}_j = z - z_j$ are shifted coordinates, and
$\chi_j$ are arbitrary coefficients.
The values $x_j$ and $z_j$ characterize the initial positions 
of fundamental solitons, 
but the actual beam trajectories may not follow the specified points 
due to mutual interactions between fundamental solitons.
Each fundamental soliton is characterized by an eigenvalue
$k_j = r_j + i \mu_j$.
Its real part, $r_j$, determines the amplitude of the fundamental
soliton, while the imaginary part, $\mu_j = \tan \theta_j$, accounts for
the soliton velocity (i.e. motion in transverse direction).
Here $\theta_j$ is the angle of the fundamental soliton propagation
relative to the $z$ axis.

To distinguish coherent and incoherent contributions to the multi-soliton
complex, we use variables $n_j$, which represent the number of the component
where the $j$-th soliton is located. Thus, two fundamental solitons with
$n_j = n_m$ are coherent, and they are incoherent otherwise. Now
we can write the expression for the matrix $D$:
\begin{equation} \label{eq:Djm}
  D_{jm} = \frac{ e_j e_m^* }{k_j + k_m^*} +
       \left\{ {
          \begin{array}{ll}
            1 / {\left( k_j + k_m^* \right)} \;,&  n_j = n_m \;, \\
            0 \;, & n_j \neq n_m \;.
          \end{array}
       } \right.
\end{equation}

Finally, the $N$-soliton solution of the original Eq.~(\ref{eq:A})
can be obtained by adding up of all the $u_j$ corresponding to a given
component number $m$:
\begin{equation}
     \psi_m = \sum_{j;\; n_j=m} u_j / \sqrt{\alpha_m}.
\end{equation}
Note that the number of terms in the sum is exactly the number of
fundamental solitons  polarized in this component, viz. $N_m$,
and the total $N$ is \mbox{$\sum_{m=1}^M N_m$}.

One of the features of this approach is that coherent fundamental solitons
are "split" among all the $u_j$ functions for a given component.
However, when obtaining analytical solutions in explicit form, it is possible
to separate fundamental solitons by combining terms with corresponding
propagation constants. Consequently, we write the exact solutions for a
different set of functions $\widetilde{u}_j$, with each of them containing one
fundamental soliton (at distances where coherent interactions are small).
These are combined into the original functions in the following way:
$\psi_m = \sum_{j;\; n_j=m} \widetilde{u}_j / \sqrt{\alpha_m}$.

The coefficients $\chi_j$ are arbitrary, and we can choose particular values 
for them:
\begin{equation} \label{eq:chi_j}
\chi_j = \prod\limits_{m;\; n_m \neq n_j}
                   \sqrt{ b_{jm} },
\end{equation}
where $b_{jm} = (k_j + k_m^*) / (k_j - k_m)$, and
the square root value is taken on the branch with positive real part.
This step significantly simplifies further analysis, as the resulting
solution will acquire a highly symmetric form.

Finally, the explicit expressions for solutions can be found as sums
over specific permutations:
\begin{eqnarray} \label{eq:sol_u}
   \widetilde{u}_j & = & \frac{e^{i \gamma_j}}{U}
           \sum_{\{ 1, \ldots ,j-1,j+1,\ldots,N \} \rightarrow
                        {L} }
              {C_L^j F_L^j (x,z)},
           \nonumber \\ \\
    U & = &
           \sum_{\{1, \ldots ,N\} \rightarrow
                        {L}}
              {C_{L} F_{L} (x,z)}. \nonumber
\end{eqnarray}
Here $L$ denotes four sets of indices $(L_1, L_2, L_3, L_4)$.
The summation is performed over all combinations in which the given set of
soliton numbers (for example, $\{1, \ldots ,N\}$) can be
split among all the $L_j$.
When performing permutations,
$L_1$, $L_2$ are only filled with numbers of mutually coherent solitons
(thus the number of elements in these sets is the same).

The coefficients and functions from~(\ref{eq:sol_u}) are determined for each
realization of the permutation $L$ as follows:
\begin{equation}  \label{eq:FC}
{\displaystyle \begin{array}{l}
 {\displaystyle \begin{array}{ll}
       C_L  = & (-1)^{|L_1|} T_{\mathrm{sg}}^{L_1} T_{\mathrm{sg}}^{L_2}
             T_{\mathrm{mg}}
             T_{\mathrm{sb}}^{L_3} T_{\mathrm{sb}}^{L_4}
             T_{\mathrm{mb}}
             T_{\mathrm{mgb}} ,
       \end{array} } \\*[9pt]
 {\displaystyle \begin{array}{ll}
            F_L (x,z)  = &  \cos(S_{\mathrm{g}})
                     \cos(S_{\mathrm{f}})
                     \cosh(S_{\mathrm{b}}) - \\
       &              \sin(S_{\mathrm{g}})
                       \sin(S_{\mathrm{f}})
                       \sinh(S_{\mathrm{b}}) ,
       \end{array} } \\*[9pt]
 {\displaystyle \begin{array}{ll}
     C_L^j   = & (-1)^{|L_1|} T_{\mathrm{c}}^j
             T_{\mathrm{sg}}^{L_1} T_{\mathrm{sg}}^{L_2}
             T_{\mathrm{mg}}   T_{\mathrm{g}}^j
             T_{\mathrm{sb}}^{L_3} T_{\mathrm{sb}}^{L_4}
             T_{\mathrm{mb}}
             T_{\mathrm{mgb}} ,
       \end{array} } \\*[9pt]
 {\displaystyle \begin{array}{ll}
   F_L^j (x,z)   = &       \cos(S_{\mathrm{g}}^j)
                       \cos(S_{\mathrm{f}})
                       \cosh(S_{\mathrm{b}}^j) - \\
       &               \sin(S_{\mathrm{g}}^j)
                        \sin(S_{\mathrm{f}})
                        \sinh(S_{\mathrm{b}}^j) .
       \end{array} } 
  \end{array} }
\end{equation}
Here we used $|L_l|$ to denote the number of elements in the set.
Note that the $F$ functions are
written in the simplest form in terms of trigonometric and hyperbolic
functions, due to the specific choice of coefficients in
Eq.~(\ref{eq:chi_j}).

The variables introduced above are the following sums and products over
the $L_j$ sets:
\begin{eqnarray*}
&& T_{\mathrm{c}}^j = \left[ 1 +
                      \sum_{m \in L_1;\; n_m = n_j} 1 \right]^{-1} , \\
  && T_{\mathrm{sg}}^{L_l}  =
        \prod_{ \{j,m\} \in L_l;\; j<m} {\left\{{ \begin{array}{ll}
                     | k_j - k_m |^2, & n_j = n_m , \\
                     s_{jm} |k_j + k_m^*|, &   n_j \neq n_m ,
                      \end{array} }\right.} \\
  && T_{\mathrm{mg}} =
        \prod_{ j \in L_1;\; m \in L_2} {\left\{{ \begin{array}{ll}
                     1 / | k_j + k_m^* |^2 , & n_j = n_m ,\\
                     s_{jm} / |k_j - k_m| , &   n_j \neq n_m ,
                      \end{array} }\right.} \\
  && T_{\mathrm{g}}^{j} =
        \prod_{ m \in L_1 \cup L_2 \cup L_3 \cup L_4}
              {\left\{{ \begin{array}{ll}
                     1 / c_{jm} ,      &   n_j = n_m , \\
                     s_{jm} \sqrt{ c_{jm} } , &   n_j \neq n_m ,
                      \end{array} }\right.} \\
  && T_{\mathrm{sb}}^{L_l}  =
        \prod_{ \{j,m\} \in L_l;\; j \leq m} {\left\{{ \begin{array}{ll}
                     1 / (2 r_j) , & j = m , \\
                     c_{jm}^{-2} , & n_j = n_m , \\
                     1  ,          &  n_j \neq n_m ,
                      \end{array} }\right.} \\
  && T_{\mathrm{mb}} =
        \prod_{ j \in L_3;\; m \in L_4} {\left\{{ \begin{array}{ll}
                     1 ,      & n_j = n_m , \\
                     c_{jm} , & n_j \neq n_m ,
                      \end{array} }\right.} \\
  && T_{\mathrm{mgb}}  =
        \prod_{ \begin{array}{l}
                 m_1 \in L_1 \cup L_2 \\
                 m_2 \in L_3 \cup L_4 \end{array} }
                {\left\{{ \begin{array}{ll}
                     1 / c_{{m_1}{m_2}} ,    & n_{m_1} = n_{m_2} , \\
             s_{{m_1}{m_2}} \sqrt{c_{{m_1}{m_2}}} , & n_{m_1} \neq n_{m_2} ,
                      \end{array} }\right.} \\
  && S_{\mathrm{g}} =
        \sum_{ j \in L_l} \gamma_j - \sum_{ j \in L_2} \gamma_j ,\,
   S_{b} =  \sum_{ j \in L_3} \beta_j - \sum_{ j \in L_4} \beta_j , \\
  && S_{\mathrm{g}}^j = S_{\mathrm{g}} -
        i \left( S_{\mathrm{sg}}^{j,L_1}-S_{\mathrm{sg}}^{j,L_2}\right),\\
  && S_{\mathrm{sg}}^{j,L_l} =
        \sum_{ m \in L_l} {\left\{{ \begin{array}{ll}
                     2 \eta_{jm} , & n_j = n_m , \\
                     \eta_{jm} ,   & n_j \neq n_m ,
                      \end{array} }\right.} \\
  && S_{\mathrm{b}}^j = S_{\mathrm{b}} +
        i \left( S_{\mathrm{sb}}^{j,L_3}-S_{\mathrm{sb}}^{j,L_4}\right) , \\
  && S_{\mathrm{sb}}^{j,L_l} =
        \sum_{ m \in L_l} {\left\{{ \begin{array}{ll}
                     2 \varphi_{jm} , & n_j = n_m , \\
                     \varphi_{jm} ,   & n_j \neq n_m ,
                      \end{array} }\right.} \\
  && S_{\mathrm{f}} = S_{\varphi}^{L_1, L_3} + S_{\varphi}^{L_2, L_4}
             - S_{\varphi}^{L_1, L_4} - S_{\varphi}^{L_2, L_3} , \\
  && S_{\varphi}^{L_{l1}, L_{l2}} =
        \sum_{ j \in L_{l1};\; m \in L_{l2}} {\left\{{ \begin{array}{ll}
                     2 \varphi_{jm} , & n_j = n_m , \\
                     \varphi_{jm} ,   & n_j \neq n_m .
                      \end{array} }\right.}
\end{eqnarray*}
Here the "$\cup$" operator is used to merge the sets, and the variables
$\beta_j + i \gamma_j = k_j \bar{x}_j + i k_j^2 \bar{z}_j / 2$
(with $\beta_j$ and $\gamma_j$ real),
$\eta_{jm} = \log \left( \left|(k_j - k_m) (k_j + k_m^* ) \right| \right)/2$,
$c_{jm} = | b_{jm}|$,
$\varphi_{jm} = \arg {\left( 1 / b_{jm} \right)} / 2$,
$s_{jm} = {\mathrm{sign}}\left\{ \pi - {\mathrm{arg}}\left[
  \sqrt{ b_{jm}} \left( \sqrt{ b_{mj} } \right)^*
  / {\left( k_j + k_m^* \right)} \right] \right\}
 {\mathrm{sign}}\left( m - j \right)$,
$b_{jm} = (k_j + k_m^*) / (k_j - k_m)$.
The function $\arg$ is supposed to give values in the interval
$\left[ 0,\; 2 \pi \right)$, and
\[
  {\mathrm{sign}} = \left\{ \begin{array}{ll}
           1, & x \geq 0 , \\
          -1, & x < 0 .
        \end{array} \right.
\]
Note that only $\beta_j$ and $\gamma_j$ depend on the coordinates
$(x,z)$. All the other coefficients are expressed in terms of the wave
numbers $k_j$ and constant shifts in positions $(x_j,\; z_j)$ of
the $N$ fundamental solitons. As the total solution has
translational symmetry, one of the shifts can be fixed, so that the number of
independent parameters controlling the multisoliton complex is $2N-1$.

If an incoherent soliton consists only of orthogonally polarized fundamental
solitons ($n_j \equiv j$, $N \equiv M$), and all are
propagating in the same direction,
then its transverse intensity profile remains stationary \cite{Ank99}. 
In this particular case, the general expressions (\ref{eq:FC}) 
are radically simplified, since,
due to the above-mentioned restrictions on the permutations, 
the sets $L_1$ and $L_2$ are always empty.
Hence, we obtain: 
\begin{eqnarray*}
  && C_L = T_{\mathrm{mb}},\; C_L^j = 2 r_j \chi_j T_{\mathrm{mb}} , \\
  && F_L = \cosh(S_{\mathrm{b}}),\; F_L^j = \cosh(S_{\mathrm{b}}^j) .
\end{eqnarray*}
Note that here we have neglected a common multiplier in $C_L$ and $C_L^j$, 
as these coefficients determine respectively the denominator and numerator
in the expression for $\widetilde{u}_j$.

\begin{figure}
\setlength{\epsfxsize}{5cm}
\setlength{\epsfysize}{5.5cm}
\centerline{\mbox{\epsffile{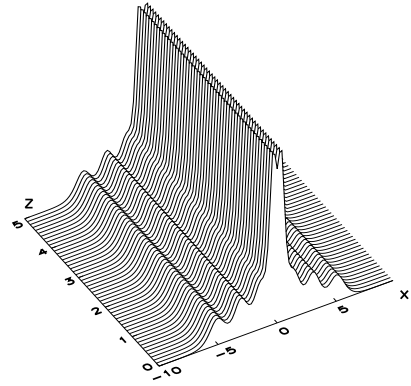}}}
\vspace{5mm}
\caption{\label{fig:prof_incoh}
Stationary propagation of an incoherent soliton consisting of
eight completely incoherent fundamental solitons 
(polarized in different components).}
\end{figure}

Now we present numerical examples to illustrate these results.
An example of a stationary incoherent soliton consisting of eight 
components ($N=M=8$) is shown in Fig.~\ref{fig:prof_incoh}. 
The profiles of the constituent fundamental
solitons, and their superposition as a whole, are determined by the
wave numbers and relative shifts along the $x$ axis.
In this configuration, the shifts in propagation direction, $z_j$, 
correspond to arbitrary phase changes of different components,
but these do not influence the evolution due to the incoherent nature of
the inter-component interactions.

\begin{figure}
\setlength{\epsfxsize}{5cm}
\setlength{\epsfysize}{5cm}
\centerline{\mbox{\epsffile{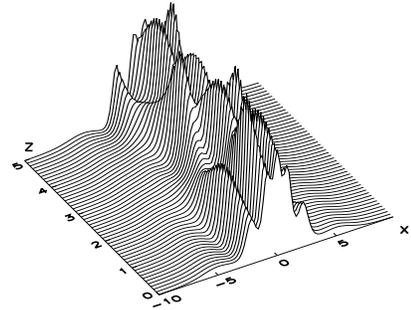}}}
\vspace{5mm}
\caption{\label{fig:prof_mix}
 Evolution of an incoherent soliton with multi-scale periodic
 "beating" due to internal coherent interactions (8 fundamental
 solitons in 5 components).}
\end{figure}

On the other hand, if $N>M$, two or more of the fundamental solitons
are polarized in the same components, and thus interact coherently.
If the inclination angles of the fundamental solitons are all the same, 
the beam will remain localized upon propagation. Such a multi-soliton complex 
is an incoherent soliton with an intensity profile which
evolves periodically or quasi-periodically, as shown in
Fig.~\ref{fig:prof_mix}. 
These oscillations, appearing due to
internal coherent intra-component interactions, are a general feature of
incoherent solitons, and can be eliminated only in specific cases, 
as discussed earlier.
It follows that spatial "beating" always accompanies the interaction of
fundamental solitons of a single NLSE, which agrees with previous studies
\cite{book}.

Our explicit solution (\ref{eq:sol_u})
also describes collisions of incoherent solitons.
As mentioned earlier, the polarizations of the fundamental solitons are
preserved in collisions (provided they are orthogonal or parallel),
and thus the degree of internal coherence doesn't change. 
However, the shifts of the fundamental soliton trajectories differ,
and this results in the incoherent solitons changing their shapes. 
These transformations can be seen clearly in Fig.~\ref{fig:prof_collis}.

\begin{figure}
\vspace{-2.5cm}
\setlength{\epsfxsize}{6cm}
 \centerline{\mbox{\epsffile{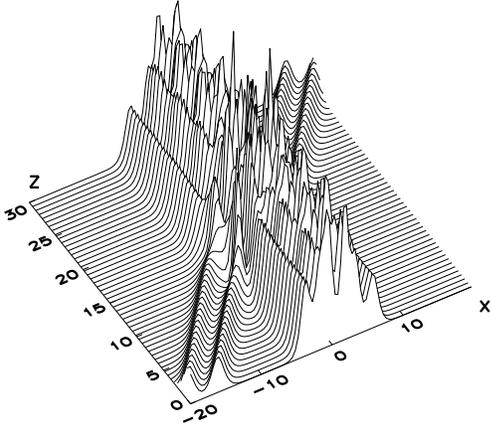}}}
 \vspace{5mm}
 \caption{ \label{fig:prof_collis}
   Collision of a completely incoherent soliton
   (consisting of two orthogonally polarized fundamental solitons)
   and an incoherent soliton with internal coherent contributions
   (6 fundamental solitons in 5 components).}
\end{figure}

To calculate the shifts, we use the fact that in the expression for
soliton profiles, $\widetilde{u}_j$, given by Eq.~(\ref{eq:sol_u}),
the denominator $U$ is real, and
the numerator does not depend on the coordinates 
of the corresponding fundamental soliton $(x_j,z_j)$. 
It is then straightforward to take appropriate limits and 
calculate the shift of \mbox{$j$-th} fundamental
soliton along the $x$ axis due to collisions:
\[
   \delta x_j = \frac{1}{r_j} \sum_m \pm
           {\left\{ \begin{array}{ll}
                  2 \ln( c_{jm}) , & n_j = n_m , \\
                  \ln( c_{jm}) ,   & n_j \neq n_m .
                     \end{array} \right.}
\]
Here the summation involves the fundamental solitons which
feature in the collisions. The "$+$" sign corresponds to the case when
colliding soliton number $m$ comes from the right (i.e. has larger $x$
coordinate before the impact), and the "$-$" sign when from the left. 
This is a generalization of the expressions found in 
\cite{Ank99}.

In summary, we have obtained a general $N$-soliton solution of $M$ coupled
nonlinear Schr\"odinger equations which describes
multi-soliton complexes supported by a Kerr-type nonlinearity.
A particular example is an incoherent soliton in a photo-refractive medium.
We have revealed that the properties of multi-soliton complexes, which are
superpositions of fundamental solitons with orthogonal or parallel
polarizations, are determined by internal interactions, both
phase-insensitive inter-component and coherent intra-component, with the
latter resulting in spatial "beating".
Using our exact result, we also analyzed collisions of incoherent solitons.
We showed that the re-shaping of incoherent solitons after collisions are
characterized by the relative shifts of the fundamental solitons, 
and these are calculated using a simple analytical formula.
These distinctive features of incoherent solitons are illustrated by
numerical examples.

The authors are part of the Australian Photonics CRC.
We are grateful to Dr. Ankiewicz for critical reading of this manuscript.

\end{narrowtext}
\end{multicols}

\begin{references}
\bibitem{Mitchell_Segev}
M. Mitchell and M. Segev, Nature (London) 387, 880 (1997).
\bibitem{Incohfirst}
M. Mitchell, Z. Chen, M. Shih, and M. Segev,
Phys. Rev. Lett. {\bf 77}, 490 (1996).
\bibitem{newstuff}
M. I. Carvalho, T. H. Coskun, D. N. Christodoulides, M. Mitchell, and M. Segev,
Phys. Rev. E {\bf 59}, 1193 (1999).
\bibitem{pr_solitons1}
G. Duree {\it et al}, Phys. Rev. Lett. {\bf 71}, 533 (1993);
\bibitem{pr_solitons6}
A. A. Zozulya, D. Z. Anderson, A. V. Mamaev, and M. Saffman, Europhys.
Lett. {\bf 36},  419 (1996).
\bibitem{pr_solitons2}
M. D. Iturbe-Castillo {\it et al}, Appl. Phys. Lett. {\bf 64}, 408 (1994)
\bibitem{Vysl2}
V. Kutuzov,  V. M. Petnikova, V. V. Shuvalov, and V. A. Vysloukh,
Phys. Rev. E {\bf 57}, 6056 (1998).
\bibitem{Hasegawa1}
A. Hasegawa,
Physics of Fluids {\bf 20}, 2155 (1977).
\bibitem{Bashkin}
E. P. Bashkin and A. V. Vagov,
Phys. Rev. B {\bf 56}, 6207 (1997).
\bibitem{Hasegawa2}
A. Hasegawa,
Opt. Lett. {\bf 5}, 416 (1980).
\bibitem{Makh1} V. G. Makhan'kov and O. K. Pashaev,
Teor. Mat. Fiz. {\bf 53}, 55 (1982)
[Theor. Math. Phys. {\bf 53}, 979 (1982)].
\bibitem{ZS}
V. E. Zakharov and A. B. Shabat,
Zh. Eksp. Teor. Fiz. {\bf 61}, 118 (1971)
[Sov. Phys. JETP {\bf 34}, 62 (1971)].
\bibitem{Manakov}
C. B. Manakov,
Zh. Teor. Eksp. Fiz. {\bf 65}, 505 (1973)
[Sov. Phys. JETP {\bf 38}, 248 (1974)].
\bibitem{PCS}
N. N. Akhmediev, W. Kr\'{o}likowski, and A. W. Snyder,
Phys. Rev. Lett. {\bf 81}, 4632 (1998).
\bibitem{Nogami}
Y. Nogami and C. S. Warke,
Phys. Lett. {\bf 59A}, 251 (1976).
\bibitem{Kaw75}
I. Kay and H. E. Moses,
J. Appl. Phys. {\bf 27}, 1503 (1956).
\bibitem{Ank99}
A. Ankiewicz, W. Kr\'{o}likowski, and N. N. Akhmediev,
Phys. Rev. E (to be published in May, 1999).
\bibitem{Gardner}
C. S. Gardner, J. M. Greene, M. D. Kruskal, and R. M. Miura,
Comm. Pure Appl. Math. {\bf 27}, 97 (1974).
\bibitem{book}
N.~Akhmediev and A. Ankiewicz,
{\it Solitons, nonlinear pulses and beams}
(Chapman \& Hall, 1997).

\end{references}
\end{document}